\documentclass{llncs}
\usepackage{llncsdoc}

\usepackage{graphicx}
\usepackage{bm}
\usepackage{latexsym}
\usepackage{times}
\usepackage{amssymb}
\newenvironment{mytheorem}{\begin{theorem}}{\end{theorem}}

\newcommand{\ie}{{\em i.e.,}\ }
\newcommand{\eg}{{\em e.g.,}\ }

\newenvironment{myproof}{\noindent {\bf\em  Proof:}\\}{$\blacksquare$\vspace{12pt}}
\def\cornerbox#1#2#3{\setbox1=\hbox{#1}
\dp1=0pt\par\hangindent\wd1\hangafter-#2\noindent
\hskip-\wd1\raise#3\box1\ignorespaces}

\begin{document}
\bibliographystyle{splncs}
\mainmatter
\title{A Generalization of Tsallis' Non-Extensive Entropy\\
and Energy Landscape Transformation Functions}
\author{Mark Fleischer}
\institute{Johns Hopkins University\\
Applied Physics Laboratory\\Laurel, Maryland
20723\\\email{Mark.Fleischer@jhuapl.edu}}

\date{\today}

\maketitle
\begin{abstract}
This article extends the non-extensive entropy of Tsallis and uses
this entropy to model an energy producing system in an absorbing
heat bath.  This modified non-extensive entropy is superficially
identical to the one proposed by Tsallis, but also incorporates a
``hidden'' parameter that provides greater flexibility for
modeling energy constraints.  This modified non-extensive entropy
also leads to a more generalized family of energy transformation
functions and also exhibits the structural scale invariance
properties described in a previous article.  This energy
transformation also provides a more natural mechanism by which
arbitrary power-law distributions can be stated in exponential
form.
\end{abstract}

\section{Introduction}\label{sec:intro}
\font\huge=ptmr at 42pt \cornerbox{\vtop{\kern 1.5pt\hbox{\huge
I\kern 2pt}}}{3}{6pt}n a recent article, Fleischer
\cite{Fleischer-SISRNSM} described a number of scale invariant and
symmetry properties in non-extensive systems---systems based on a
non-extensive form of entropy developed by Tsallis
\cite{Tsallis88}.  These scale invariant properties show that in
aggregating energy states of a large ensemble, many of the
mathematical relationships associated with a given state also
apply to the aggregations and use identical mathematical forms.
Investigation of this scale invariance property revealed an {\em
energy transformation function}.  The frequency of its appearance
in these scale invariant mathematical forms suggests that energy
transformations are an important element in modeling non-extensive
systems. This energy landscape transformation also provides an
additional perspective on non-extensivity.  Tsallis' entropic form
essentially shifts the probability weights in the system and
thereby creating a power law distribution.  This same effect can
be captured by the energy landscape transformation functions.
Thus, the energy landscape transformation function apparently lies
at the heart of the scale invariance and symmetry properties
present in non-extensive systems and seems to account for many of
the useful applications of the Tsallis entropy especially in
modeling dissipative systems where energy producing systems in a
heat bath lose energy to its environment.

In this article, new forms of scale invariance and energy
transformations are illuminated by generalizing (extending)
Tsallis' entropy formulation to explicitly account for and
generalize this energy transformation property. In this vein,
Tsallis' entropy and all of its associated scale invariance and
symmetry properties become special cases in a larger array of
scale invariance and symmetry. These new properties show a number
of features involving recursion (aggregations of aggregations) and
ways of characterizing power-law distributions using exponential
forms where the exponent is the transformed energy value.  In this
sense, this article expands on the results in
\cite{Fleischer-SISRNSM}.

This article is organized as follows: Section~\ref{sec:background}
provides some background on Tsallis entropy and the scale
invariant properties described earlier.
Section~\ref{sec:modifiedTsallis} describes the new form of
entropy and related system constraints.  These two components are
then used to define a new stationary probability.
Section~\ref{sec:SI-Su} briefly describes the scale invariance
properties based on the forms of the new entropy and stationary
probability and the associated {\em family} of energy
transformation functions. Section \ref{sec:symmetry} explores the
related symmetry and power law relationships.  Finally,
Section~\ref{sec:discussion} offers some discussion on the
implications of this modified Tsallis entropy, energy
transformations and practical applications in the field of complex
system simulation.  Section \ref{sec:conclusion} provides some
concluding remarks.

\section{Background}\label{sec:background}
Tsallis \cite{Tsallis03,Tsallis88} developed a new entropy
expression that forms the basis of a {\em non-extensive} form of
thermodynamics:
\begin{equation}\label{eqn:Sq}
S_q = \frac{k\left(1-\sum_{i=1}^Wp_i^q\right)}{q-1}
\end{equation}
where $k$ is a constant and $S_q$ is the entropy parameterized by
the entropic parameter $q$.  In classical statistical mechanics,
entropy falls into a class of variables that are referred to as
{\em extensive} because they scale with the size of the system.
{\em Intensive} variables, such as temperature, do not scale with
the size of the system.\footnote{Combine two vessels of gas each
with the same volume and pressure into another vessel of twice the
volume and the pressure and temperature of the combined gas will
be the same as before.  Energy and entropy, however, are examples
of {\em extensive} variables in that combining several sources of
either energy or entropy and you increase the total energy or
entropy.}  Tsallis' form of entropy is {\em non-extensive} because
the entropy of the union of two independent systems is not equal
to the sum of the entropies of each system. That is, for
independent systems $A$ and $B$,
\begin{equation}
S_q(A+B) = S_q(A)+S_q(B) +
\frac{(1-q)S_q(A)S_q(B)}{k}.\label{eqn:Sq(A+B)}
\end{equation}
Tsallis uses this entropy to calculate a stationary probability
$p_i$ for a canonical system where $S_q$ is maximized. Tsallis
shows that $p_i$ is distributed by a power law
\begin{equation}\label{eqn:def-p}
p_i(t) = \frac{\left[1 +
\left(\frac{q-1}{t}\right)f_i\right]^{\frac{1}{1-q}}}{Z_q} \ =\
\frac{\left[1 + a f_i\right]^{\frac{1}{1-q}}}{Z_q}
\end{equation}
where $a=(q-1)/t$ and based on system constraints (but see
\cite{Fleischer-SISRNSM,Curado91} for a discussion on constraints)
\begin{eqnarray}
\sum_{i=1}^Wp_i& = & 1\label{eqn:sumpi=1}\\
\sum_{i=1}^Wp^q_if_i &=& U, \mbox{ a
constant.}\label{eqn:sumpqf=U}
\end{eqnarray}
This distribution is different than the exponential law in the
classic Boltzmann-Gibbs distribution, symbolized here by $\pi_i$.
Tsallis points out that for $q\rightarrow 1$ the extensivity
properties of classical statistical mechanics emerge. Thus, \eg
$\lim_{q\rightarrow 1}S_q=S_{BG} = -k\sum_i\pi_i\ln\pi_i$, and
$\lim_{q\rightarrow 1}p_i(t) = \pi_i(t)$, hence the Tsallis
entropy is a generalization of the Boltzmann-Gibbs entropy
\cite{Tsallis88,Tsallis03}.

Tsallis \cite{Tsallis04} also notes that the form of
(\ref{eqn:Sq}) is the simplest form that satisfies certain
non-additivity assumptions and uses only one parameter, namely the
entropic parameter $q$ .  In the sections that follow, we both
retain this inherent simplicity, but at the same time incorporate
a new, ``hidden'' parameter $m$ that provides additional
flexibility for modeling dissipative systems.

\subsection{Scale Invariance and Symmetry in Extensive and Non-Extensive
Systems}\label{sec:SIbackground} In
\cite{Fleischer-SISRNSM,FleischerSAscale02} a number of scale
invariant properties in both classical and non-extensive systems
are described and are based on aggregations of energy states.
Briefly, this type of structural scale invariance is best
illustrated by the following: for any aggregated set of energy
levels $A = \{i_1, i_2,\ldots,i_n\}$ where index $i_k$ refers to
some particular energy level (note $A$ can also simply consist of
a single energy energy level $i$),
\begin{equation}\label{eqn:derivpiA}
\frac{\partial \pi_A(t)}{\partial t} =
\frac{\pi_A(t)}{t^2}\left[f_A(t) - \langle f\rangle(t)\right]
\end{equation}
where
\[
\pi_A(t) = \sum_{i\in A}\pi_i(t) \mbox{\ \ and\ \ } f_A(t)=
\frac{\sum_{i\in A}\pi_i(t)f_i}{\pi_A(t)}
\]
with the latter equivalent to the {\em conditional expectation} of
a random energy value $f$ given that the current state $i$ is in
set $A$. A similar form of scale invariance exists for second
moments. Thus,
\begin{equation}\label{eqn:deriv<fA>}
\frac{\partial f_A(t)}{\partial t} = \frac{\sigma^2_A(t)}{t^2}
\end{equation}
where $\sigma_A^2(t)$ is the conditional variance of energy
(objective function) values at temperature $t$ given that the
current state being in set $A$. See
\cite[p.232-33]{FleischerSAscale02} for details and a more formal
treatment.

Fleischer \cite{Fleischer-SISRNSM} demonstrated a similar result
for non-extensive systems where the only difference between the
scale invariance properties of the classic case and that of the
non-extensive case was that the latter involved an {\em energy
transformation function}. Thus, for any aggregated set of energy
levels $A$,
\begin{equation}\label{eqn:derivpA}
\frac{\partial p_A(t)}{\partial t} =
\frac{p_A(t)}{t^2}\left[\hat{f}_A(t) - \langle
\hat{f}\rangle(t)\right]
\end{equation}
where
\begin{equation}\label{eqn:def-aggregated-p-f} p_A(t) =
\sum_{i\in A}p_i(t) \mbox{\ \ and\ \ } \hat{f}_A(t)=
\frac{\sum_{i\in A}p_i(t)\hat{f}_i(t)}{p_A(t)}
\end{equation}
where the transformed energy value $\hat{f}_i$ is defined by
\begin{equation}
\hat{f}_i(f_i,q,t) \equiv \frac{f_i}{1 +
\left(\frac{q-1}{t}\right)f_i} = \frac{f_i}{1 + a\,f_i}
\end{equation}
where $a = (q-1)/t$.  It will henceforth be notationally
convenient to simply refer to transformed energy value using
$\hat{f}_i(t),\hat{f}_i$ or $\hat{f}_A$ as the case may be where
it is understood to depend on the parameters $i,f_i,q,t$.

Fleischer \cite{Fleischer-SISRNSM} also showed that the scale
invariance in second moments incorporates this energy
transformation, hence for all aggregations $A$
\begin{eqnarray}
\frac{\partial \hat{f}_A}{\partial t}&=& \frac{{\langle
\hat{f}^2\rangle}_A - {\langle\hat{f}\rangle}_A^2 +(q-1){\langle
\hat{f}^2\rangle}_A}{t^2}\nonumber\\
&=&\frac{\hat{\sigma}^2_A}{t^2}+ \frac{(q-1){\langle
\hat{f}^2\rangle}_A}{t^2}.\label{eqn:derivfhatAend}
\end{eqnarray}

\subsection{Discussion}
The appearance of the transformed energy value $\hat{f}$ in these
scale invariant forms suggests that it is an important component
in non-extensive statistical mechanics. It is therefore reasonable
to infer that for an energy absorbing heat bath, the
``equilibrium'' condition associated with the internal energy
constraint in (\ref{eqn:sumpqf=U}) implicitly involves this
transformed energy value.  One viewpoint suggests that in an
energy producing system in thermal equilibrium with an energy
absorbing heat bath there must be some way of characterizing the
rate at which the energy is produced and absorbed. To generalize
this notion further, a heat bath that absorbs energy at a high
rate, must be balanced by a higher rate (in some sense) of energy
production {\em if one seeks to model some equilibrium condition}.
We can then perhaps model different rates of energy production and
absorption by making the absorption rate proportional to some
function of the energy level in a canonical system.  In this way,
the notion of non-extensivity can be further expanded and modeled
in a fashion that encompasses the information/entropy loss, as
indicated in (\ref{eqn:Sq(A+B)}) for $q>1$, and an equilibrium
condition that encompasses an energy production/absorption
component.

To capture these notions in a more flexible way, a modified
Tsallis entropy $S_u$ is defined and in a manner that provides
additional flexibility for modeling canonical systems in energy
absorbing heat baths yet retains the inherent simplicity of the
non-extensive entropy. The increased flexibility is based on using
a parameter $u$ that involves both the entropic parameter $q$ from
Tsallis' entropy and an energy parameter $m$. Note that in
Tsallis' approach, the entropic parameter $q$ is typically
associated with the state probability $p_i$ (see \eg
(\ref{eqn:Sq}) and (\ref{eqn:sumpi=1})). Tsallis notes the effects
of the exponent $q$ in the Type 2 constraint as they
``[privilegiate] the {\em rare} and the {\em frequent\/} events''
depending on whether $q<1$ or $q>1$, respectively
\cite[p.535]{Tsallis98}.  But this notion of shifting the
probability weight of different energy values is, in some sense,
equivalent to transforming the energy values themselves.  Thus,
the entropic parameter $q$ plays a central role, in capturing this
shift of probability by transforming the energy landscape as
indicated in the derivation of the energy transformation function
in (\ref{eqn:def-fhat}). Thus, $q > 1$ implies that the exponent
of $f_i$ in (\ref{eqn:sumpqf=U}) is also 1.  It would however be
useful to permit some additional freedom in choosing the exponent
of $f_i$ to capture this notion of an energy producing system in
equilibrium with an absorbing heat bath.

Taking these considerations into account, the parameter $u$ which
involves both $q$ and $m$ should be associated with the state
probability $p_i$, and so will also affect the energy
transformation function.  Because we also want to fashion
situations where the exponent of $f_i$ in (\ref{eqn:sumpqf=U}) is
not always equal to 1, we require that the exponent of $f_i$ in
(\ref{eqn:sumpqf=U}) be different from that of $p_i$. Keeping
these qualifications in mind, the following section describes a
modified non-extensvie entropy $S_u$ where a number of new scale
invariant properties emerge.

\section{The Modified Non-Extensive Entropy $S_u$}\label{sec:modifiedTsallis}
In this section, Tsallis' entropy is
generalized to increase the flexibility in using energy
transformation functions.  In this formulation, the Tsallis
entropic parameter $q$ in (\ref{eqn:Sq}) is replaced by an
entropic/energy parameter $u$ ($u$ always follows $q$) that
simultaneously accounts for both the parameter $q$ and an energy
transformation parameter $m\geq 1$ where
\begin{equation}\label{eqn:def-u}
u = (q-1)(m-1)+1
\end{equation}
and the {\em modified Tsallis entropy} is defined by
\begin{equation}\label{eqn:Su}
S_u \equiv \frac{k\left(1-\sum_{i=1}^Wp_i^u\right)}{u-1}.
\end{equation}
Note that this form is identical to (\ref{eqn:Sq}), hence retains
all of its inherent benefits.  It is easy to see that
\begin{equation}\label{eqn:limitchainSu}
\lim_{u\rightarrow 1}S_u\ =\ \lim_{q\rightarrow 1}S_u\ =\
\lim_{q\rightarrow 1}S_q\ =\ S_1 = S_{\mbox{\tiny BG}} \mbox{
(defined earlier)}
\end{equation}
and certain other relationships and properties are easily extended
and generalized. For example, (\ref{eqn:Sq(A+B)}) becomes
\begin{eqnarray}
S_u(A+B)& = &S_u(A)+S_u(B) + \frac{(1-u)S_u(A)S_u(B)}{k}\nonumber \\
& = &S_u(A)+S_u(B) +
\frac{(1-q)(m-1)S_u(A)S_u(B)}{k}\label{eqn:Su(A+B)}
\end{eqnarray}
since $u-1=(q-1)(m-1)$.  Thus, for any given value of $q>1$ where
some entropy loss occurs, the magnitude of this loss can also be
modeled using the parameter $m$ which, as explained below, is
associated with the energy levels. Note that if $q=1$, the value
of $m$ becomes irrelevant. The next section explores the
implications of this simple modification.

\subsection{The Energy Loss Rate}
The parameter $m$ is useful for modeling an {\em energy producing
system} in thermal equilibrium with an absorbing heat bath.  Such
a system dissipates its energy to its surroundings while
maintaining an average value of its internal energy.  A canonical
ensemble of such a system can be modeled therefore by a {\em
power} of the energy function $f$ as in
\begin{equation}\label{eqn:energyconstraint}
\sum_{i=1}^W p_i^uf_i^{\,m-1} = U \mbox{ a constant}
\end{equation}
where the parameter $m$ serves to capture the notion of an {\em
energy loss rate} or energy dissipation rate (or perhaps, the
energy absorption rate). The exponent $u$ of the stationary
probability $p_i$ thus serves the same purpose as in Tsallis'
works (see \cite{Curado91,Tsallis98}). Notice that $(m=2)
\Rightarrow (u=q)$, and the resulting system is equivalent to
those based strictly on $S_q$.  Notice also that for $m=2$, this
constraint has the same form as that described by Tsallis' Type 2
constraint in (\ref{eqn:sumpqf=U}) and also in Tsallis' Type 3
constraint involving the so-called ``escort probabilities'' (see
\cite{Tsallis98}).

\subsection{The Stationary Probability}
Tsallis \cite{Tsallis88,Tsallis98} illustrates how maximizing the
entropy $S_q$ given the constraints in (\ref{eqn:sumpi=1}) and
(\ref{eqn:sumpqf=U}) leads to a stationary probability and results
in the well-known stationary probability in (\ref{eqn:def-p}). To
obtain the stationary probability $p_i$ in light of the modified
Tsallis entropy $S_u$ subject to the normalization constraint in
(\ref{eqn:sumpi=1}) and the internal energy constraint in
(\ref{eqn:energyconstraint}) we use a similar approach as in
\cite{Tsallis88,Curado91}.  The general Lagrangian function is
\begin{equation}
L = k\left(\frac{1-\sum_{i=1}^Wp_i^u}{u-1}\right) +
\alpha\left(\sum_{i=1}^Wp_i(t)-1\right) - \beta\left(\sum_{i=1}^W
p^u_if_i^{\,m-1} - U\right).\label{eqn:def-L}
\end{equation}
Therefore,
\begin{equation}\label{eqn:partialL}
\frac{\partial L}{\partial p_i}= \frac{up_i^{u-1}}{1-u}  + \alpha
- \beta up_i^{u-1}f_i^{\,m-1}.
\end{equation}
Setting (\ref{eqn:partialL}) to zero and rearranging, we obtain
\begin{displaymath}
\frac{up_i^{u-1}}{u-1}\left[1+\beta (u-1)f_i^{\,m-1}\right] =
\alpha
\end{displaymath}
and hence
\begin{eqnarray}
p_i
&=&\left[\frac{\alpha(u-1)}{u}\right]^{\frac{1}{u-1}}\left[1+\beta
(u-1)f_i^{\,m-1}\right]^{\frac{-1}{u-1}}\nonumber\\
&=&\frac{\left[1+\beta
(u-1)f_i^{\,m-1}\right]^{\frac{1}{1-u}}}{Z_u}\label{eqn:def_pinu}
\end{eqnarray}
where $\beta$ is often symbolized by the inverse temperature $1/t$
and
\[
Z_u = \sum_i\left[1+\beta
(u-1)f_i^{\,m-1}\right]^{\frac{1}{(1-u)}}
\]
is the corresponding normalization constant. Note that
(\ref{eqn:def_pinu}) has the same form as (\ref{eqn:def-p}) except
for the exponent of $f_i$. Noting that $u-1 = (q-1)(m-1)$ we
obtain the general form
\[
p_i = \frac{\left[1+\beta
(q-1)(m-1)f_i^{\,m-1}\right]^{\frac{1}{(1-q)(m-1)}}}{Z_u}.
\]
Letting $a = \beta(q-1)=(q-1)/t$ for notational convenience, then
\begin{equation}\label{eqn:def-ponSu}
p_i =
\frac{\left[1+a(m-1)f_i^{\,m-1}\right]^{\frac{1}{(1-q)(m-1)}}}{Z_u}.
\end{equation}
Notice that for the case $m=2$, (\ref{eqn:def-ponSu}) is
equivalent to (\ref{eqn:def-p}).

\section{SA Scale Invariance Based on $S_u$}
\label{sec:SI-Su}

To demonstrate scale invariance based on aggregated states in the
non-extensive case involving $S_u$, we proceed in a similar
fashion as in \cite{Fleischer-SISRNSM} while making the necessary
adjustments to account for the definition of $u$.  The following
theorem states a scale invariant mathematical structure for
systems based on $S_u$.
\begin{mytheorem}\label{thm:scaleinvariance}
Let $A=\{i_1,i_2,\ldots, i_n\}$ be any aggregation of energy
levels $i$ where the energy level associated with $i$ is denoted
as $f_i$.  Using the definition of $p_i$ in (\ref{eqn:def-ponSu}),
define the stationary probability of set $A$ by $p_A = \sum_{i\in
A}p_i$ and energy value associated with aggregated sets $A$ by
\[
\hat{f}^{\,m-1}_A = \frac{\sum_{i\in A}p_i\hat{f}_i^{\,m-1}}{p_A}
\]
where
\[
\hat{f}^{\,m-1}_i \equiv \frac{f_i^{\,m-1}}{1 +
a(m-1)f^{\,m-1}_i}.
\]
Define the conditional variance of $\hat{f}^{\,m-1}_A$ by
\[
\hat{\sigma}^2_A = \langle \hat{f}^{2m-2}\rangle_A - \langle
\hat{f}^{m-1}\rangle_A^2.
\]
Then
\[
\frac{\partial p_A}{\partial t}=
\frac{p_A}{t^2}\left[\hat{f}^{\,m-1}_A - \langle \hat{f}^{\,m-1}
\rangle\right]
\]
and
\[
\frac{\partial \hat{f}^{\,m-1}_A}{\partial t}=
\frac{\hat{\sigma}^2_A}{t^2}+
\frac{(u-1)\langle\hat{f}^{2m-2}\rangle_A}{t^2}.
\]
\end{mytheorem}

\begin{myproof}
It is convenient to first consider the relevant quantities
associated with individual states (energy levels) $i$, which in
this case refers to $\partial p_i(t)/\partial t$ and
$\partial\langle \hat{f}^{\,m-1}_i \rangle/\partial t$ where these
quantities are defined below.  We proceed in similar fashion as in
\cite{Fleischer-SISRNSM,FleischerSAscale02} taking into account
the definition of $u,S_u$ and the exponent of $f_i$.

\noindent For notational convenience and simplicity, let $N_i(t)$
be the numerator in (\ref{eqn:def-ponSu}). Thus,
\begin{equation}\label{eqn:def-Ni}
N_i(t)\equiv
\left[1+\left(\frac{u-1}{t}\right)f^{\,m-1}_i\right]^{\frac{1}{1-u}}
\end{equation}
(hereinafter we will drop the $(t)$ from $N_i(t)$ to further
simplify the expressions) and taking the derivative of
(\ref{eqn:def-ponSu}) with respect to temperature $t$,
\begin{equation}\label{eqn:deriv-p-withN}
\frac{\partial p_i(t)}{\partial t} =
\frac{Z_u\frac{\partial}{\partial t}N_i -
N_i\frac{\partial}{\partial t}Z_u}{(Z_u)^2}.
\end{equation}
In this case,
\begin{equation}
\frac{\partial  N_i}{\partial t} = \frac{\partial}{\partial
t}\left[1 + \left(\frac{u-1}{t} \right)f_i^{\,m-1}
\right]^{\frac{1}{1-u}}= \frac{N_i^uf^{\,m-1}_i}{t^2},
\label{eqn:derivNi}
\end{equation}
with $Z_u = \sum_jN_j,$ and $\  p_i(t)=N_i/Z_u$ and hence
\begin{displaymath}
 \frac{\partial}{\partial t}Z_u =
 \frac{\partial}{\partial t}\sum_jN_j = \sum_j\frac{\partial
N_j}{\partial t} =\sum_j\frac{N_j^uf^{\,m-1}_j}{t^2}.
\end{displaymath}
Substituting this and (\ref{eqn:derivNi}) into
(\ref{eqn:deriv-p-withN}) yields
\begin{eqnarray}
\frac{\partial p_i(t)}{\partial t} &=&
\frac{\left(\sum_jN_j\right)\frac{N^u_if^{\,m-1}_i}{t^2} -
\frac{N_i}{t^2}\sum_jN_j^uf^{\,m-1}_j}{(\sum_jN_j)^2}\nonumber\\
&=& \frac{p_i(t)N_i^{u-1}f^{\,m-1}_i}{t^2}-
\frac{p_i(t)\sum_jN_j^uf^{\,m-1}_j}{t^2\sum_jN_j}\nonumber\\
&=&\frac{p_i(t)}{t^2}\left[N_i^{u-1}f^{\,m-1}_i-\frac{\sum_jN_j^uf^{\,m-1}_j}{\sum_jN_j}\right]\label{eqn:deriv-p-b4fhatdef}
\end{eqnarray}
To further simplify the notation, define the {\em transformed
energy value}
\begin{eqnarray}
\hat{f}^{\,m-1}_i &\equiv& N^{\,u-1}_if^{\,m-1}_i\nonumber\\
&=&\frac{f^{\,m-1}_i}{1+a(m-1)f^{\,m-1}_i} =
\frac{f^{\,m-1}_i}{M_{i,m}}\label{eqn:def-fhat}
\end{eqnarray}
where $a=(q-1)/t$ and,
\begin{equation}\label{eqn:def-Mi}
M_i = 1 + a(m-1)f^{\,m-1}_i
\end{equation}
(the reason for this latter definition will become clear later on)
without the clutter of the arguments. Substituting
(\ref{eqn:def-fhat}) into (\ref{eqn:deriv-p-b4fhatdef}) and
further
simplifying yields the two equivalent forms:
\begin{eqnarray}
\frac{\partial p_i}{\partial t}
&=&\frac{p_i}{t^2}\left[\hat{f}^{\,m-1}_i- \langle
\hat{f}^{\,m-1}\rangle\right]\label{eqn:deriv-p-enda}\\
&=&\frac{p_i}{t^2}\left[\frac{f^{m-1}_i}{M_i} - \left\langle
\frac{f^{m-1}}{M}\right\rangle\right].\label{eqn:deriv-p-endb}
\end{eqnarray}

\noindent Now, taking the derivative of the probability of the
aggregated set,
\begin{equation}\label{eqn:deriv-lumped-1}
\frac{\partial p_A}{\partial t} = \frac{\partial }{\partial
t}\sum_{i\in A}p_i = \sum_{i\in A}\frac{\partial p_i}{\partial t}.
\end{equation}
and substituting (\ref{eqn:deriv-p-enda}) into
(\ref{eqn:deriv-lumped-1}) (keeping in mind the dependence on $t$)
yields
\begin{eqnarray}
\frac{\partial p_A}{\partial t} &=& \sum_{i\in
A}\frac{p_i}{t^2}\left[\hat{f}^{\,m-1}_i- \langle
\hat{f}^{\,m-1}\rangle\right]\nonumber\\
&=&\sum_{i\in A}\frac{p_i\hat{f}^{\,m-1}_i}{t^2} -\sum_{i\in
A}\frac{p_i\langle\hat{f}^{\,m-1}\rangle}{t^2}\nonumber\\
&=&\frac{p_A}{t^2}\sum_{i\in A}\frac{p_i\hat{f}^{\,m-1}_i}{p_A}-
\frac{p_A\langle\hat{f}^{\,m-1}\rangle}{t^2}\nonumber\\
&=&\frac{p_A}{t^2}\left[\hat{f}^{\,m-1}_A -
\langle\hat{f}^{\,m-1}\rangle\right]\label{eqn:deriv-pA}
\end{eqnarray}
where for aggregated states, (\ref{eqn:deriv-pA}) has a similar
mathematical structure as
(\ref{eqn:deriv-p-enda},\ref{eqn:deriv-p-endb}), hence exhibits a
scale invariance property the foundation of which is based on the
energy transformation function $\hat{f}^{\,m-1}_i$.

Scale invariance for second moments is indicated in the following
where again the energy transformation of $f_i$ is used. Keeping in
mind the dependence of $\hat{f}$ and $p_i$ on the temperature $t$,
consider
\begin{eqnarray}
\frac{\partial \langle \hat{f}^{\,m-1} \rangle}{\partial t}& =&
\frac{\partial \hat{f}^{\,m-1}_\Omega}{\partial t}=
\frac{\partial}{\partial t} \left[
\sum_{i\in\Omega}p_i\hat{f}^{\,m-1}_i\right]\nonumber\\
&=&\sum_{i\in\Omega}\frac{\partial}{\partial
t}\left[p_i\hat{f}^{\,m-1}_i\right]\nonumber\\
&=&\sum_{i\in\Omega}\left[\frac{\partial p_i}{\partial
t}\hat{f}^{\,m-1}_i + p_i\frac{\partial
\hat{f}^{\,m-1}_i}{\partial
t}\right].\label{eqn:deriv-<fhat>Omega}
\end{eqnarray}
Substituting (\ref{eqn:deriv-p-enda}) into the first part of
(\ref{eqn:deriv-<fhat>Omega}) and simplifying yields
\begin{equation}
\frac{\partial \langle \hat{f}^{\,m-1} \rangle}{\partial t}=
\sum_{i\in\Omega}\frac{p_i\hat{f}^{\,2m-2}_i}{t^2}-
\sum_{i\in\Omega}\frac{\hat{f}^{\,m-1}_\Omega
p_i\hat{f}^{\,m-1}_i}{t^2}+
\sum_{i\in\Omega}p_i\frac{\partial\hat{f}^{\,m-1}_i}{\partial
t}\label{eqn:threetermsofderiv-<fhat>Omega}.
\end{equation}
Noting the form of the first two terms on the right-hand-side in
(\ref{eqn:threetermsofderiv-<fhat>Omega}) and the fact that in the
third term
\begin{equation}\label{eqn:derivhatf}
\frac{\partial \hat{f}^{\,m-1}_i}{\partial t} =
\hat{f}^{\,2m-2}_i\left(\frac{u-1}{t^2}\right)
\end{equation}
and substituting into (\ref{eqn:threetermsofderiv-<fhat>Omega})
and adding the symbol $\Omega$ to denote expectations over the
entire state space yields
\begin{eqnarray}
\frac{\partial {\langle \hat{f}^{\,m-1} \rangle}_\Omega}{\partial
t} &=& \frac{{\langle
\hat{f}^{\,2m-2}\rangle}_\Omega-{\langle\hat{f}^{\,m-1}\rangle}_\Omega^2+(u-1){\langle
\hat{f}^{\,2m-2}\rangle}_\Omega}{t^2}\nonumber\\
&=&\frac{\hat{\sigma}_\Omega^2}{t^2} +
\frac{(u-1){\langle\hat{f}^{\,2m-2}\rangle}_\Omega}{t^2}\label{eqn:deriv-fhat-Omega}
\end{eqnarray}
where $\hat{\sigma}_\Omega^2$ represents the variance of the
values $\hat{f}^{\,m-1}_i$ over the entire energy landscape (at
temperature $t$).

Eq. (\ref{eqn:deriv-fhat-Omega}) provides the basis for another
form of scale invariance.  Thus, after going through similar steps
as in (\ref{eqn:deriv-<fhat>Omega}) through
(\ref{eqn:deriv-fhat-Omega}) we get
\begin{equation}\label{eqn:derivfhatAraw}
\frac{\partial \hat{f}^{\,m-1}_A}{\partial t}= \frac{\sum_{i\in
A}p_i \hat{f}^{\,2m-2}_i}{t^2p_A} -
\frac{\hat{f}^{\,2m-2}_A}{t^2}+ \left(\frac{u-1}{t^2}
\right)\frac{\sum_{i\in A}p_i\hat{f}^{\,2m-2}_i}{p_A}.
\end{equation}
Noting that the first and third terms indicate conditional
expectations conditioned on the current state being in set $A$,
then (\ref{eqn:derivfhatAraw}) can be re-written in the convenient
notation
\begin{eqnarray}
\frac{\partial \hat{f}^{\,m-1}_A}{\partial t}&=& \frac{{\langle
\hat{f}^{\,2m-2}\rangle}_A - {\langle\hat{f}^{\,m-1}\rangle}_A^2
+(u-1){\langle
\hat{f}^{\,2m-2}\rangle}_A}{t^2}\nonumber\\
&=&\frac{\hat{\sigma}^2_A}{t^2}+ \frac{(u-1){\langle
\hat{f}^{\,2m-2}\rangle}_A}{t^2}\label{eqn:derivfhatAend}
\end{eqnarray}
where (\ref{eqn:derivfhatAend}) is clearly analogous to
(\ref{eqn:deriv-fhat-Omega}) and so exhibits a form of scale
invariance.  Note that the terms involving $(u-1)$ also scale with
the aggregated set $A$ in the non-extensive case.  See
\cite{Fleischer-SISRNSM,FleischerSAscale02} for similar results in
the case of $S_q$ and $S_1$, respectively.
\end{myproof}

An interesting aspect of these relationships can be succinctly
described using the following commutative-like diagram in
Figure~\ref{fig:commute}. A similar diagram can be produced to
depict an analogous relationship for second-moments.  The next
section further explores some of these symmetry relationships.

\def\harr#1#2{\smash{\mathop{\hbox to
.75in{\rightarrowfill}}\limits^{\displaystyle#1}_{\displaystyle
#2}}}

\def\varr#1#2{\llap{$\scriptstyle #1$}\left\downarrow\vcenter to .75in{}\right.\rlap{$\scriptstyle #2$}}

\def\diagram#1{{\normallineskip=8pt
\normalbaselineskip=0pt \matrix{#1}}}
\begin{figure}[h]
\vspace{-.6in}
\[
\diagram{(p)&\harr{q=1}{}&(\pi)\cr
\varr{\textstyle\frac{\partial}{\partial
t}}{}&&\varr{}{\textstyle\frac{\partial}{\partial t}}\cr
\left(\frac{\partial p}{\partial
t}\right)&\harr{q=1}{}&\left(\frac{\partial \pi}{\partial
t}\right)\cr}
\]
\begin{center}\parbox[t]{3.25in}{\vspace{-.25in}\caption{Symmetry in the
sequence of operations: The limit of the derivative equals the
derivative of the limit.}\label{fig:commute}}
\end{center}
\end{figure}

\section{Symmetry Relationships}\label{sec:symmetry}
Fleischer \cite{Fleischer-SISRNSM} explored a number of symmetry
relationships in addition to ones Tsallis has indicated in his
early works based on the effects of different constraints used in
modeling a canonical system (see also \cite{Tsallis98}). In this
section, we briefly identify the analogous symmetries described in
\cite{Fleischer-SISRNSM} in light of the modified non-extensive
entropy $S_u$. The reader is referred to \cite{Fleischer-SISRNSM}
for background.

In Tsallis' original incarnation of non-extensive statistical
mechanics, he used his ``Type 1'' constraint $\sum_i\tilde{p}_if_i
= U$ (see \cite{Tsallis98}) where the exponent $q$ as in $p^q_i$
was absent. This led to the following expression for the
stationary probability
\begin{equation}\label{eqn:def-ptilde}
\tilde{p}_i = \frac{\left[1 + \left(\frac{1-q}{t}\right)f_i
\right]^{\frac{1}{q-1}}}{Z_q^\prime}
\end{equation}
where $\tilde{Z}_q^\prime$ is the obvious normalizing constant.
Tsallis notes that the form of $\tilde{p}_i$ is essentially the
same as that of $p_i$ except that $1-q$ replaces every occurrence
of $q-1$ and vice versa {\em including in the exponents}.  Here,
we simply note that solving the corresponding Lagrangian function
in (\ref{eqn:def-L}) using the constraint
\begin{equation}\label{eqn:SuType1}
\sum_i\tilde{p}_if^{\,m-1}_i = U
\end{equation}
leads, not surprisingly, to the analogous equation
\begin{equation}\label{eqn:def-ptildeSu}
\tilde{p}_i(t) = \frac{\left[1 -a(m-1)f^{\,m-1}_i
\right]^{\frac{1}{(q-1)(m-1)}}}{\tilde{Z}_u}
\end{equation}
which, since $a = (q-1)/t$, has the same reversal in sign observed
in \cite{Fleischer-SISRNSM}.

Fleischer \cite{Fleischer-SISRNSM} further showed that systems
with the constraint (\ref{eqn:SuType1}) also entail a similar
scale invariance as previously highlighted except that the
corresponding energy transformation function is also modified with
a sign change.  In the context of $S_u$ and the constraint in
(\ref{eqn:SuType1}), the energy transformation function that is
the basis of the scale invariance is given by
\begin{equation}\label{eqn:def-tildef-Su}
\tilde{f}^{\,m-1}_i = \frac{f^{\,m-1}_i}{1 - a(m-1)f^{\,m-1}_i}
\end{equation}
where $a = (q-1)/t$ as before. Again, the analogous sign change in
the denominator of (\ref{eqn:def-tildef-Su}) versus the
denominator in (\ref{eqn:def-fhat}) is present.

\subsection{Scale Invariance Using Other Constraints}
Using the same approach as in \cite{Fleischer-SISRNSM} and in
Section~\ref{sec:SI-Su}, it follows that for all aggregated energy
levels $A$, and using the constraint in (\ref{eqn:SuType1}) leads
to the scale invariant form
\begin{equation}
\frac{\partial \tilde{p}_A}{\partial t} =
\frac{\tilde{p}_A}{t^2}\left[\tilde{f}^{\,m-1}_A -
\langle\tilde{f}^{\,m-1}\rangle\right]\label{eqn:deriv-tildepA-Su}
\end{equation}
where again the only difference between this and the earlier
result is that every occurrence of $p_i$ and $\hat{f}_i$ is
replaced with a $\tilde{p}_i$ and $\tilde{f}_i$, respectively.

\subsubsection{Second Moments}
Proceeding in the same fashion as in (\ref{eqn:deriv-<fhat>Omega})
through (\ref{eqn:derivfhatAend}) and using analogous definitions
(\ie $\tilde{\sigma}^2$ corresponds to the variance of the values
of $\tilde{f}^{\,m-1}_i$) we obtain the result
\begin{eqnarray}
\frac{\partial {\langle \tilde{f}^{\,m-1} \rangle}}{\partial t}
&=& \frac{{\langle
\tilde{f}^{2m-2}\rangle}-{\langle\tilde{f}^{\,m-1}\rangle}^2+(1-u){\langle
\tilde{f}^{2m-2}\rangle}}{t^2}\nonumber\\
&=&\frac{\tilde{\sigma}_\Omega^2}{t^2} +
\frac{(1-u){\langle\tilde{f}^{2m-2}\rangle}}{t^2}.\label{eqn:derivftildeOmega}
\end{eqnarray}
(the symbol $\Omega$ serves as a reminder that these values are
based on the variation over the entire landscape).  Scale
invariance in second moments with the Tsallis Type 1 constraint is
indicated by
\begin{eqnarray}
\frac{\partial \tilde{f}^{\,m-1}_A}{\partial t}&=& \frac{{\langle
\tilde{f}^{2m-2}\rangle}_A - {\langle\tilde{f}^{\,m-1}\rangle}_A^2
+(1-u){\langle
\tilde{f}^{\,2m-2}\rangle}_A}{t^2}\nonumber\\
&=&\frac{\tilde{\sigma}^2_A}{t^2}+ \frac{(1-u){\langle
\tilde{f}^{\,2m-2}\rangle}_A}{t^2}\label{eqn:derivftildeAend}
\end{eqnarray}
where (\ref{eqn:derivftildeOmega}) and (\ref{eqn:derivftildeAend})
are similar to (\ref{eqn:deriv-fhat-Omega}) and
(\ref{eqn:derivfhatAend}) except that, as before, every occurrence
of $u-1$ and $\hat{f}$ has been replaced with a $1-u$ and
$\tilde{f}$, respectively.

\subsection{Probabilities and Energy Relations}
A number of additional forms of symmetry relating to the
constraints are described in \cite{Fleischer-SISRNSM} and their
forms in the context of $S_u$ are easily inferred.  Thus, \eg
defining $f_i$ in terms of $\hat{f}_i$ we have
\begin{equation}\label{eqn:ffromfhat}
f_i = \frac{\hat{f}_i}{\left[1 -
a(m-1)\hat{f}^{\,m-1}_i\right]^{\frac{1}{m-1}}}
\end{equation}
where again we use $a = \left(\frac{q-1}{t}\right)$ for notational
convenience.  Using this we can define the probability $p_i$ in
terms of $\hat{f}_i$ and obtain
\begin{equation}
p_i(t) = \frac{\left[1 -
a(m-1)\hat{f}^{\,m-1}_i\right]^{\frac{1}{(q-1)(m-1)}}}{Z_u}\label{eqn:1-over-q-1}
\end{equation}
which as in \cite{Fleischer-SISRNSM} is exactly the same form as
in (\ref{eqn:def-ptilde}) above where the Type 1 constraint was
used except with the $\hat{f}_i$ replacing the $f_i$. It follows
therefore that
\begin{equation}
Z_u = \sum_{i=1}^W\left[1 + a(m-1)f^{\,m-1}_i
\right]^{\frac{1}{(1-q)(m-1)}}\ =\ \sum_{i=1}^W\left[1
-a(m-1)\hat{f}^{\,m-1}_i \right]^{\frac{1}{(q-1)(m-1)}}.
\end{equation}

The following lemma expands on Lemma 1 in
\cite{Fleischer-SISRNSM}.
\begin{lemma}
For all $u > 1$,
\begin{equation}
\sum_ip^u_if^{\,m-1}_i = Z_u^{1-u}\sum_ip_i\hat{f}^{\,m-1}_i =
Z_u^{1-u}\langle \hat{f}^{\,m-1} \rangle.
\end{equation}
\end{lemma}
\begin{myproof}
It follows from the definition of $p_i$, that for all $i$,
\begin{equation}\label{eqn:pqlemmaproof}
p^u_if^{\,m-1}_i = \left(\frac{\left[1 + \left(\frac{u-1}{t}
\right)
f^{\,m-1}_i\right]^{\frac{1}{1-u}}}{Z_u}\right)^uf^{\,m-1}_i.
\end{equation}
Now observe that $\frac{u}{1-u} = \frac{1}{1-u} - 1$. Consequently
for all $i$,
\begin{eqnarray*}
p_i^u &=& \frac{\left[1 + \left(\frac{u-1}{t} \right)
f^{\,m-1}_i\right]^{\frac{1}{1-u}}}{Z_uZ_u^{u-1}\left[1 +
\left(\frac{u-1}{t} \right) f^{\,m-1}_i\right]}\\
&=& \frac{p_i}{Z_u^{u-1}\left[1 + \left(\frac{u-1}{t} \right)
f^{\,m-1}_i\right]}.
\end{eqnarray*}
Substituting this into (\ref{eqn:pqlemmaproof}) and simplifying we
get
\begin{equation}
p_i^uf^{\,m-1}_i = Z_u^{1-u}\,\,p_i\hat{f}^{\,m-1}_i
\end{equation}
and summing over all $i$ the result follows.
\end{myproof}

We further note that other analogous symmetries as in
\cite{Fleischer-SISRNSM} are present with regard to the
relationships involving the Type 1 and Type 2 constraints.  These
relationships simply highlight the significance of the energy
transformation function which the next section explores further.
The following delineates these relationships the proofs of which
correspond to those in \cite{Fleischer-SISRNSM} and so are omitted
here.

\begin{lemma}\label{lem:recursion}
For any energy index $i$ and parameters $a>0$ and $m\geq 1$,
define the $k+1^{\mbox{\tiny th}}$ iterate of the energy
transformation of $f_i$ in terms of the $k^{\mbox{\tiny th}}$
iterate by
\[
\hat{f}^{m-1}_{i,k+1} = \frac{\hat{f}^{m-1}_{i,k}}{1 +
a(m-1)\hat{f}^{\,m-1}_{i,k}}
\]
with $\hat{f}_{i,0} \equiv f_i$ and
\[ \hat{f}_{i,1} \equiv
\frac{f_i}{\left[1 + a(m-1)f^{\,m-1}_i \right]^{\frac{1}{m-1}}}.
\]
Then,
\begin{equation}\label{eqn:kthiterate}
\hat{f}_{i,k} = \frac{f_i}{\left[1 + ka(m-1)f^{\,m-1}_i
\right]^{\frac{1}{m-1}}}.
\end{equation}
\end{lemma}

\begin{myproof}
This proof is a straight forward extension of the one in
\cite{Fleischer-SISRNSM}.
\end{myproof}

\subsection{A Exponential Form of Powerlaws}
In \cite{Fleischer-SISRNSM}, Fleischer notes how the energy
landscape transformation leads to an exponential form for a
power-law distribution.  It was shown that for
\begin{equation}\label{eqn:def-xhat}
\hat{x} \equiv \frac{x^\gamma}{1+ ax^\gamma}
\end{equation}
then
\begin{equation}\label{eqn:gammapowerlaw}
e^{-\lambda \hat{x}} - C_1 \sim C_2x^{-\gamma}
\end{equation}
where $C_1$ and $C_2$ are constants.  The exponent $\gamma$ in
(\ref{eqn:def-xhat}) served to generalize the basic form of the
landscape transformation function so that the power law exponent
could itself be more general.  This $\gamma$ was a component that
was added for this specific purpose.  With the definition of $S_u$
and the related energy transformation function in
(\ref{eqn:def-fhat}), however, this ``artificial'' addition
becomes unnecessary as the exponent $m-1$ in (\ref{eqn:SuType1})
more naturally yields the general exponent of the related power
law.  Consequently, the constraint in (\ref{eqn:SuType1}) and the
attendant definition in (\ref{eqn:def-fhat}) yield a general form
for the power law using an exponential form based on first
principles.  The following theorem is therefore stated and the
proof is omitted as it again follows analogously from the one in
\cite{Fleischer-SISRNSM}.

\begin{theorem}\label{thm:asymptotics}
Let $a>0$ and $x>0$ be such that $a(m-1)x^{m-1} > 1$ and define
\[
\hat{x}=\frac{x^{m-1}}{1+a(m-1)x^{m-1}}
\]
using the energy transformation function defined earlier. Then for
all $a > 0,\lambda > 0$ and $m>1$
\[
e^{-\lambda\hat{x}}-C_1\ \sim \ C_2x^{1-m}
\]
as $x\longrightarrow\infty$ where the constants $C_1=
e^{-\lambda/a(m-1)}$ and $C_2 = a^{-2}(m-1)^{-2}\lambda\,
e^{-\lambda/a(m-1)}$.
\end{theorem}
\begin{myproof}
This proof is a straight forward extension of the one in
\cite{Fleischer-SISRNSM}. Thus, by substituting $m-1$ for every
occurrence of $\gamma$ and $a(m-1)$ for every occurrence of $a$ in
\cite{Fleischer-SISRNSM}, the result follows.
\end{myproof}

\subsection{The Derivative of $\hat{f}_i$}

Finally, another important characteristic of the energy
transformation function is worth noting and explains why we have
used the exponent $m-1$ in the previous expressions.

\begin{lemma}\label{lem:diffeq}
For any $a>0$ (this is mathematically equivalent to the condition
that $q>1$) and $m>1$,
\[
\frac{\partial \hat{f}_i}{\partial a}= \frac{\partial
\hat{f}_i}{\partial q} = -\hat{f}^m_i.
\]
\end{lemma}
\begin{myproof}
The proof is straight-forward using the chain rule.  Thus, taking
the derivative, we get \jot=10pt
\begin{eqnarray*}
\lefteqn{\frac{\partial}{\partial a}\left(\frac{f_i}{\left[1 +
a(m-1)f^{m-1}_i \right]^{\frac{1}{m-1}}}\right)}\hspace{50pt}\\
&=&\frac{-f_i\left(\frac{1}{m-1}\right)\left[1 + a(m-1)f^{m-1}_i
\right]^{\frac{1}{m-1}-1}(m-1)f^{m-1}_i }{\left[1 +
a(m-1)f^{m-1}_i \right]^{\frac{2}{m-1}}}\\
\vspace{20pt}&=&\frac{-f^m_i}{\left[1 + a(m-1)f^{m-1}_i
\right]^{\frac{2}{m-1}+ \frac{m-2}{m-1}}}\\\vspace{10pt}
&=&\frac{-f^m_i}{\left[1 + a(m-1)f^{m-1}_i
\right]^{\frac{m}{m-1}}}\ =\ -\hat{f}^m_i
\end{eqnarray*}
\end{myproof}

\noindent It is apparent from the foregoing, that the energy
transformation function in (\ref{eqn:def-fhat}) is the solution of
the differential equation in stated in Lemma \ref{lem:diffeq} with
the following boundary conditions: for all $m>1, a = 0 \Rightarrow
\hat{f}_i = f_i$.

\section{Discussion}\label{sec:discussion}
Fleischer \cite{Fleischer-SISRNSM} suggested that the scale
invariance properties associated with the aggregation of energy
states may provide additional perspectives on macroscopic
power-law behavior.  The appearance of the energy landscape
transformation in these abstract aggregations and its relevance to
power law distributions indicated in Theorems
\ref{thm:scaleinvariance} and \ref{thm:asymptotics} and the more
generalized entropy form $S_u$ suggest further interesting
connections to complex systems theory which are briefly discussed
below.

\subsection{Complex Systems as Graphs}
The foregoing scale invariant properties are based on aggregating
energy levels.  Thus, portions of the energy spectrum denoted by a
label $A$ are lumped together and considered as having an energy
value of $\hat{f}_A(t)$. The energy transformation functions also
suggest that as these ``energy levels'' aggregate they lose energy
at a rate and in proportion to the magnitude of the energy level
associated with the aggregation.  Thus, if one can make a
connection between aggregations of energy levels and the
aggregations of `systems', a number of new possibilities arise in
modeling complex systems.

This notion is based on the fact that it is the components of
systems that either contribute to or detract from the energy
spectrum of a larger, more complex system. Thus, aggregating
portions of the energy spectrum is, in some sense, equivalent to
{\em aggregating those entities that contribute} to the energy
spectrum.  Using this perspective, the energy landscape
transformation functions suggests that these energy-contributing
entities {\em themselves undergo some sort of transformation}!
Thus, the energy spectrum associated with a complex system can be
modeled as a discretized energy spectrum and done in such a way
that each such aggregation has energy values disjoint from other
such aggregations. Depending on the level of discretization, it
may even be possible for each such aggregation to have {\em
nearly} identical energy spectrums yet not have any energy levels
in common. Consequently, one can define subsystems of a larger
complex system as nodes in a graph where each node possesses an
energy spectrum. Arcs can then be used to relate these subsystems
to each other in any convenient and useful way.

Modeling complex systems using graph theoretic means is of course
not new (see \cite{Albert02} for a fuller account).  But the scale
invariant properties described above seem to facilitate a more
direct connection between statistical mechanical relationships and
general modeling approaches for complex systems.  For instance, it
would be interesting to examine the implications of this
aggregation concept in the context of Ising spin glass models
where there are aggregations of lattice points or apply it in the
context of Markov random fields.

Another interesting application is in modeling systems in a
far-from-equilibrium condition, something readily accomplished
using the power law aspects associate with the energy
transformation functions.   In systems that are
far-from-equilibrium, the `local temperature' is not constant but
``fluctuating on a relatively large time scale or spatial scale
\cite[p.220]{Sornette04}.''  But the notion of temperature for
general, complex systems is somewhat problematic. How does one
define `temperature' in systems that do not involve large
ensembles of particles or ``thermodynamic'' behavior yet exhibits
something akin to being in a far-from-equilibrium condition? See
\cite[Ch.7]{Sornette04} for a complete discussion on the concept
of `temperature' in statistical mechanics.

One recent approach for dealing with these issues has been through
the use of {\em superstatistics}.  Sornette describes this
approach:
\begin{quote}
A particular class of more general statistics relevant for
nonequilibrium systems, containing Tsallis statistics [] as a
special case, has been termed `superstatistics' [] A
superstatistics arises out of the superposition of two statistics,
namely one described by ordinary Boltzmann factors $e^{-\beta E}$
and another one given by the probability distribution of $\beta$
\cite[p.220]{Sornette04}.
\end{quote}
We believe this concept of `superstatistics' is readily captured
through the use of the foregoing energy landscape transformations
and application Lemma \ref{lem:recursion} that concerns some
recursive properties of the energy transformation functions.

These recursive relationships, first indicated in
\cite{Fleischer-SISRNSM} and generalized here, show that in terms
of the energy landscape, there is a certain equivalence between
values of $q$, the number of times $k$ a system or subsystem
``undergoes'' an energy transformation, and the temperature $t$.
Recall, that the energy transformation comes from an expression
involving the aggregation of energy values.  But the recursive
aspects of described in Lemma \ref{lem:recursion} suggests that a
larger number of transformations and hence aggregations is, in
some sense, equivalent to higher values of $q$ and/or lower values
of $t$ and vice versa.  For example, given a particular $f_i$ (or
for that matter, $f_A$), a three-fold transformation
$\hat{f}_{i,3}$ with $q=2, m=2$ and $t=1$ yields
\[
\hat{f}_{i,3}(q=2,t=1) = \frac{f_i}{1+3f_i},
\]
which is equivalent to the value of a single transformation when
$q = 4, m=2$, and $t=1$. When $q=2, m=2$ and $t=1/3$, however,
$\hat{f}_{i,1}$ also yields $\frac{f_i}{1+3f_i}$.  Thus,
successive aggregations at some given temperature are equivalent,
in some sense, to fewer aggregations at a lower temperature
because $\hat{f}_{i,3}(t=1) = \hat{f}_{i,1}(t=1/3)$.

\section{Conclusion}\label{sec:conclusion}
This article proposed a slight modification of Tsallis' entropy
formulation.  This modification retains the basic form of Tsallis
entropy yet provides for greater flexibility in modeling energy
constraints.  The additional parameter, $m$, can thus be used as
an exponent of the energy value in the energy constraint equation.
Doing this leads to a more generalized energy transformation
function that also retains all of the scale invariance and
symmetry properties highlighted in \cite{Fleischer-SISRNSM}.
Moreover, this more general energy transformation function yields
a more general (and natural) power law expression given in
exponential form.

A number of possible issues were also examined in interpreting
these results.  Future research examines how these energy
transformation functions can be leveraged in a number of
applications, specifically, in developing new approaches for Monte
Carlo Markov Chain simulation methodologies.  Hopefully, the
properties highlighted here will also improve our understanding of
and capability to manage large, complex systems---something that
appears to be an increasing challenge in our time. 
\begin{center}
{\normalsize\bf Acknowledgments}
\end{center}
The writing of this article was supported under the Independent
Research and Development Program by the Johns Hopkins University
Applied Physics Laboratory.  The author wishes to thank Donna
Gregg, Sue Lee and William Blackert for their support and
encouragement during the writing of this article.

\bibliography{SETNSM,FleischerGeneral}

\begin{thebibliography}{1}

\bibitem{Fleischer-SISRNSM}
Fleischer, M.:
\newblock
\newblock (Scale invariance and symmetry relationships in non-extensive
  statistical mechanics) http://www.arXiv.org/cond-mat/0501293.

\bibitem{Tsallis88}
Tsallis, C.:
\newblock Possible generalization of boltzmann-gibbs statistics.
\newblock Journal of Statistical Physics \textbf{52} (1988)  479--487

\bibitem{Tsallis03}
Tsallis, C., Baldovin, F., Cerbino, R., Pierobon, P.:
\newblock Introduction to nonextensive statistical mechanics and
  thermodynamics.
\newblock In: The Physics of Complex Systems: New Advances \& Perspectives.
  Proceedings of the 1953-2003 Jubilee ``Enrico Fermi'' International Summer
  School of Physics (2003)

\bibitem{Curado91}
Curado, E.M.F., Tsallis, C.:
\newblock Generalized statistical mechanics: Connection with thermodynamics.
\newblock J. of Physics A \textbf{24} (1991)  L69--L72 Letters to the Editor.

\bibitem{Tsallis04}
Tsallis, C.:
\newblock 1: Nonextensvie Statistical Mechanics: Construction and Physical
  Interpretation.
\newblock Santa Fe Institute Studies in the Sciences of Complexity. In:
  Nonextensive Entropy. Oxford University Press, New York, NY. (2004)  pp.1--54

\bibitem{FleischerSAscale02}
Fleischer, M., Jacobson, S.:
\newblock Scale invariance properties in the simulated annealing algorithm.
\newblock Methodology and Computing in Applied Probability \textbf{4} (2002)
  219--241

\bibitem{Tsallis98}
Tsallis, C., Mendesc, R.S., Plastino, A.:
\newblock The role of constraints within generalized nonextensive statistics.
\newblock Physica A \textbf{261} (1998)  534--554

\bibitem{Albert02}
Albert, R., Albert-L\'{a}szl\'{o}:
\newblock Statistical mechanics of complex networks.
\newblock Reviews Of Modern Physics \textbf{74} (2002)  47--97

\bibitem{Sornette04}
Sornette, D.:
\newblock Critical Phenomena in Natural Sciences: Chaos, Fractals,
  Selforganization and Disorder: Concepts and Tools. 2nd edn.
\newblock Springer Series in Synergetics. Springer-Verlag, New York (2004)

\end{thebibliography}

\end{document}